# Title: Enabling a multifunctional telecommunications fiber optic network: Ultrastable optical frequency transfer and attosecond timing in deployed multicore fiber

**Authors:** Nazanin Hoghooghi[1], Mikael Mazur[2], Nicolas Fontaine[2], Yifan Liu[1,3], Dahyeon Lee[1,3], Charles McLemore[1,3], Takuma Nakamura[1,3], Tetsuya Hayashi[4], Giammarco Di Sciullo[5], Divya Shaji[5], Antonio Mecozzi[5], Cristian Antonelli[5], and Franklyn Quinlan[1,6]

**Affiliations:**
[1]Time and Frequency Division, National Institute of Standards and Technology, Boulder, CO 80305, USA
[2] Nokia Bell Labs, Murray Hill, NJ, USA
[3]Department of Physics, University of Colorado Boulder, 440 UCB Boulder, CO 80309, USA
[4] Sumitomo Electric Industries, Ltd., 1, Taya-cho, Sakae-ku, Yokohama, Kanagawa, Japan
[5]University of L'Aquila and CNIT, 67100 L'Aquila, Italy
[6]Electrical, Computer and Energy Engineering, University of Colorado Boulder, 425 UCB Boulder, CO 80309, USA
Author e-mail address: nazanin.hoghooghi@nist.gov, franklyn.quinlan@nist.gov

**Abstract:** The telecommunications industry's deployment of billions of kilometers of optical fiber has created a vast global network that can be exploited for additional applications such as environmental sensing, quantum networking and international clock comparisons. However, for reasons such as the unidirectionality of long-haul fiber links, telecom fiber networks cannot always be adapted for important applications beyond data transmission. Fortunately, new multicore optical fibers create the opportunity for application coexistence with data traffic, creating expansive multifunctional networks. Towards that end, we propose and demonstrate the faithful transfer of ultrastable optical signals through multicore fiber in a way that is compatible with the unidirectionality of long-haul fiber optic systems, demonstrating a fractional frequency instability of $3\times10^{-19}$ at 10,000 seconds. This opens the door towards intercontinental optical clock comparisons, with applications in fundamental physics and the redefinition of the second.

## Main Text

Fiber optic cables are the arteries of modern telecommunication networks, forming a dense web around the globe and interconnecting a data-hungry world. Amazingly, a single standard optical fiber is capable of transmitting hundreds of terabits of data every second (*1*) approaching its capacity limit. Even so, ever increasing data traffic demands, driven by cloud computing and streaming services, are predicted to soon outpace fiber optic network capabilities (*2*, *3*).

In parallel to this growth in data traffic has been the growth in other applications that look to utilize the telecom industry's investments in deployed optical fiber systems. Some of these applications can directly utilize the telecom infrastructure, allowing for simultaneous use of data-carrying optical fiber interconnects. Successful examples include environmental sensing and monitoring, where light launched through the fiber is used to measure strain and temperature changes to detect seismic events (*4*), analyze automobile traffic patterns (*5*) and aid in deep-ocean research (*6*, *7*). Other applications of deployed optical fiber, on the other hand, do not easily mesh with the network infrastructure. A notable example is the transmission of ultrastable signals from optical clocks and oscillators, of utmost importance for re-definition of the SI second (*8*, *9*), tests of fundamental physics (*10*) and geodesy (*11*). Consequently, long-distance optical clock transfer over fiber has required dedicated fibers and special equipment installations along the fiber path, and therefore does not fully reside within the standard telecom network. Moreover, these restrictive infrastructure requirements are uneconomical in undersea interconnects, and comparisons of the best optical atomic clocks over intercontinental baselines have been, unfortunately, precluded (*12*).

Here we show how advanced fiber technology, created to meet increased data traffic demands, can simultaneously offer compatibility with a range of other applications, creating a new opportunity for the formation of a vast multi-purpose network. Multicore fibers (MCF), where several light-guiding cores reside within a single fiber strand, can increase data carrying capacity to Pb/s rates on a single fiber (*13*). The prospect of MCF as the future long-haul telecom fiber has motivated deployment of test beds in Italy (*14*) and China (*15*). Moreover, MCF and associated technologies are sufficiently developed to enable commercial deployment of a subsea two-core MCF in the near future (*16*), and mass-production of MCF has begun (*17*). We add to the functionality of MCF by demonstrating state-of-the-art optical clock transfer and synchronization while operating each core unidirectionally, thus remaining compatible with standard telecom infrastructure. Over 25 km of deployed MCF, we transfer laser light with added fractional frequency instability of only $3\times10^{-19}$ at $10^4$ seconds. This added noise is low enough not to degrade the performance of the best optical atomic clocks, and can be achieved while copropagating with other, broadband light (e.g., data traffic) within the same core. Moreover, we show the relative timing error between cores to be just tens of attoseconds across 25 km of fiber, supporting synchronization at the highest level. When considering the relevance of precision timing to other applications such as quantum networking (*18–20*), it is clear that the utility of MCF goes far beyond its original goal of increasing data capacity and provides a path towards networks that serve a multiplicity of scientific and technological applications (Figure 1).

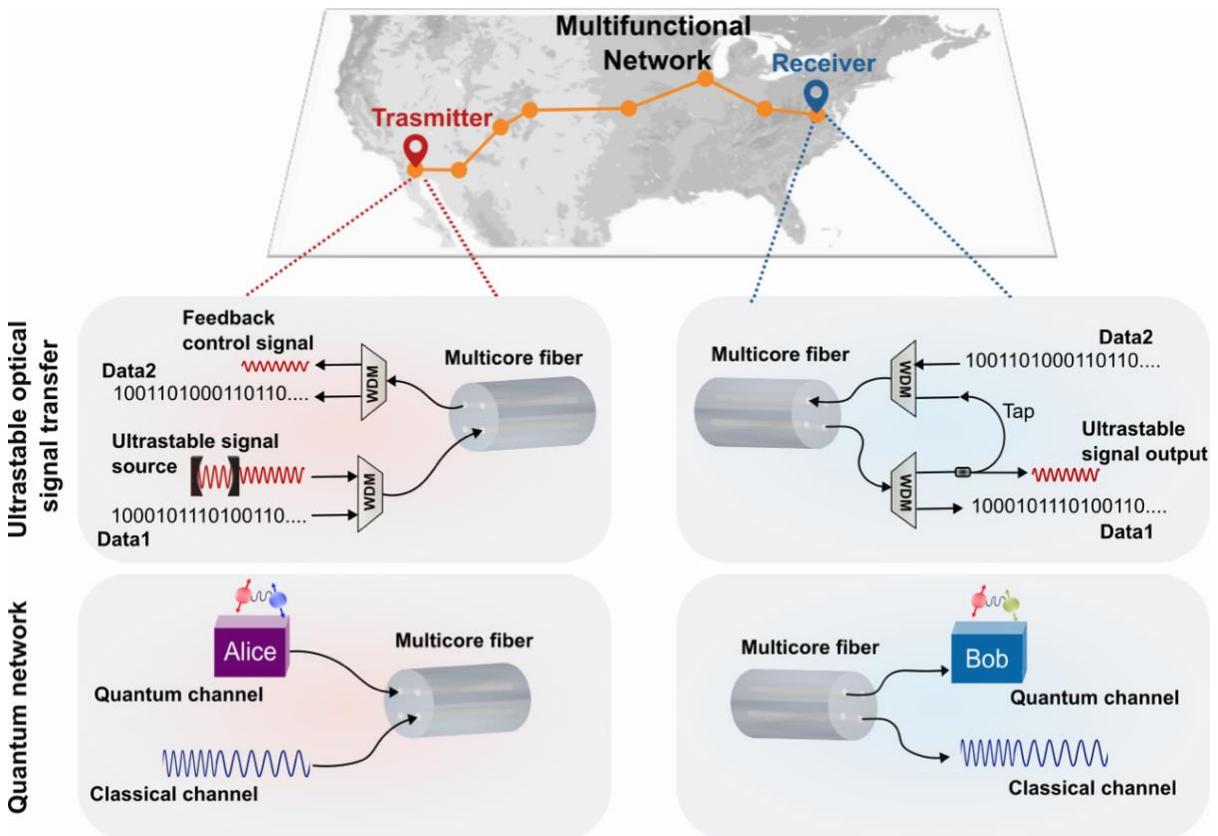

Fig. 1. **The envisioned multifunctional multicore fiber (MCF) network**. Functionalities enabled by MCF include ultra-stable optical signal transfer and coexistence of quantum and classical light in the same fiber.

The exquisite frequency stability of state-of-the-art optical clocks and oscillators, orders of magnitude more stable than their microwave frequency counterparts, cannot be transferred any significant distance without being corrupted by the transfer medium. Whereas high-fidelity ~ 100 km-scale transfer can take place over air (*21*), high performance at longer distances have all used optical fiber interconnects. With fiber, temperature, humidity and vibration all induce strain variations in the glass that couple to the optical phase and frequency of the transmitted light – a quality that is advantageous for environmental sensing but detrimental to stable frequency transfer. However, by reflecting a small portion of the light back to the transmission source, the fiber's added instability can not only be measured but compensated (*22*).

Importantly, such bidirectional use of the fiber is not supported by long-haul telecom fiber networks. Standard telecom optical amplifiers, installed at roughly 80 km intervals to compensate for fiber losses, only allow unidirectional light travel in order to reduce data transmission errors from link back-reflections and cascading noise. Hence, while optical frequency transfer has crossed 1000-kilometer distances, their need for carefully tailored bidirectional amplification has required additional hardware and signal routing that must be kept separate from the rest of the network (*23–30*) .

The adoption of MCF in long-haul networks can provide a means to integrate frequency transfer at the highest level with telecom networks. With MCF, each core within the fiber strand can be used unidirectionally but with separate cores operating with light propagating in opposite directions, including through multicore optical amplifiers (*31*). That is, one core of the MCF can be used to transmit ultrastable clock light to the end-user, while a second core is used to return light to the source for fiber noise compensation. This noise compensation scheme works as long as light along the separate cores experience the same environmentally driven phase and frequency shifts, which, as we show in more detail below, is indeed the case. Crucially, operation of MCF with different cores transmitting in different directions is not only fully compatible with telecom data transmission schemes, but is the intended configuration to reduce cross-talk among the cores. Indeed, the longest data transmission distance over MCF link to date – 18,090 km of fiber transmitting 14.9 Tb/s per core – has been achieved with this design (*32*).

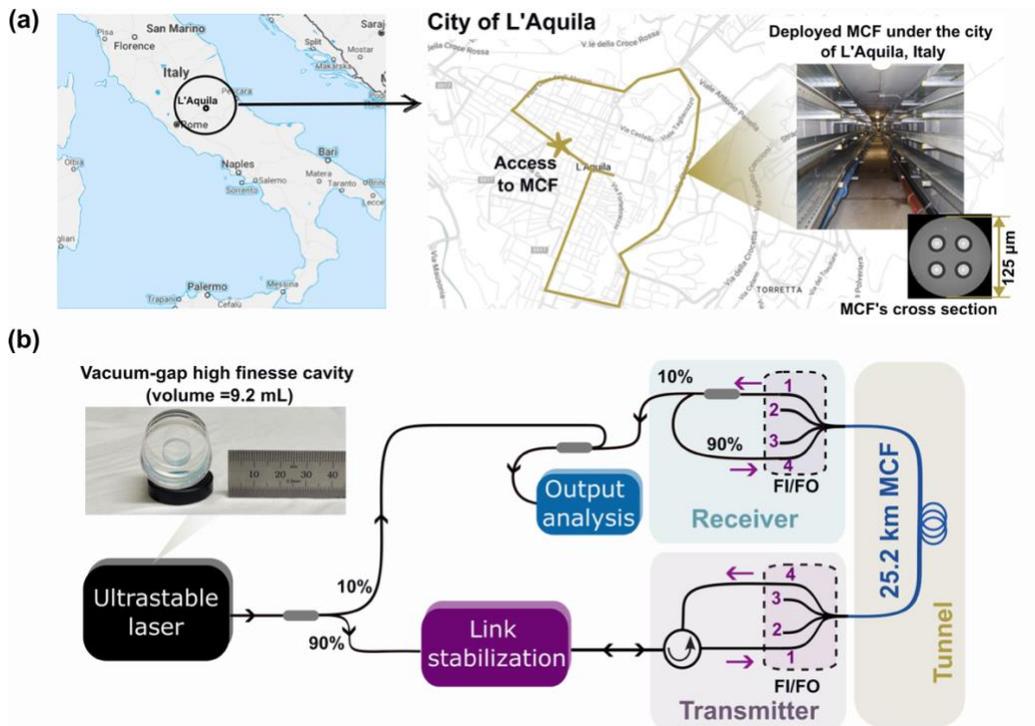

Fig. 2. **Ultrastable frequency transfer over deployed MCF**. (a)Deployed MCF under the city of L'Aquila, Italy, as part of the INCIPICT project(*33*). The 4-core MCF runs in a tunnel under the city. (b)Experimental setup of the noise-canceled deployed MCF. Four 6.3 km-long strands of MCF are cascaded, forming a 25km-long link. All-fiber components are used in this setup. A fiber circulator acts as both a beam combiner and an isolator. Light is launched into either end of the MCF link with fan-in fan-out (FI/FO) devices.

# Results

## Fractional frequency instability measurements

We demonstrate ultrastable optical frequency transfer over the world's first deployed MCF, located in the city of L'Aquila, Italy. Figure 2 shows the experimental setup. A 6.3-km MCF cable deployed in an underground tunnel network includes four strands of an uncoupled-core, 4-core fiber that can be cascaded to form a 25.2 km link (*14*). In addition to the MCF, there is a cable with 8 strands of standard fiber (SMF-28) that runs in the same tunnel under the city, providing a benchmark for our MCF stabilization performance. Both "transmission" and "receiver" ends of all the fibers are located in the same room, allowing for full evaluation of the performance of the stabilized link. As a source of ultrastable light, we use a fiber laser with a wavelength of 1550 nm whose frequency is locked to a compact optical reference cavity (*34*). Locking to the reference cavity provides laser fractional frequency stability at the $10^{-14}$ level, ensuring the link instability can be measured and compensated without corruption due to noise from the laser itself (*35*).

For performance evaluation, we split the output of the ultrastable laser, with 10% used as a reference for phase and frequency measurements of the light emanating from the remote end of the fiber. The remaining 90% of the laser light is used for transmission through the MCF and for fiber noise compensation. The fiber noise is measured by turning a round-trip path through the MCF into one arm of a laser-light interferometer, thereby converting phase and frequency fluctuations (relative to a short, free-space path in the link stabilization section) into a fluctuating interference pattern on our photodetector. Corrections are applied to laser frequency through an acousto-optic modulator (housed within the "link stabilization" section of Fig. 2), pre-compensating the laser signal before transmission through MCF.

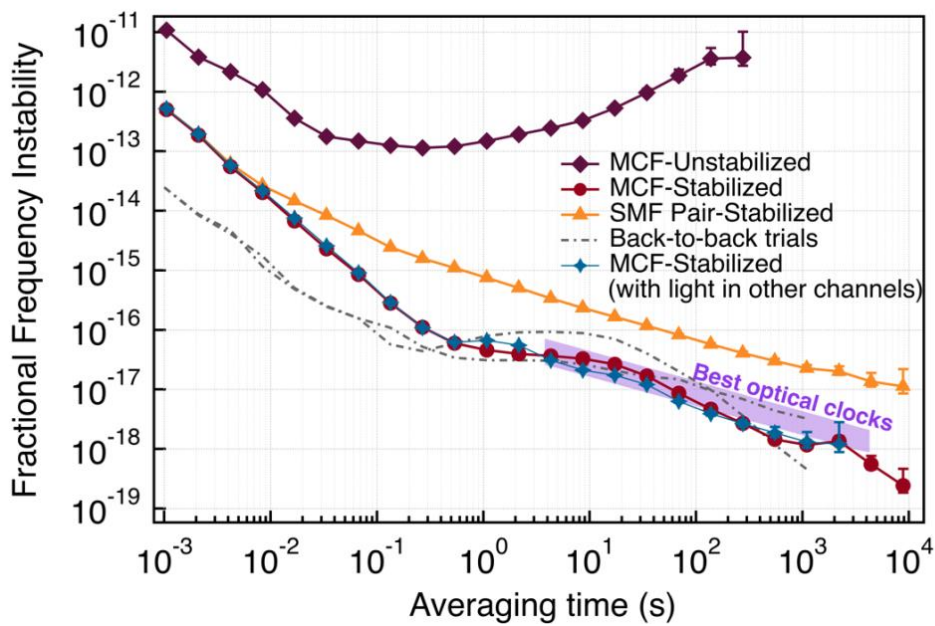

Fig. 3. **Fractional frequency instability results with the MCF.** The added instability of the stabilized and unstabilized MCF links (25.2 km), back-to-back trials (where the MCF fiber is excluded), and a stabilized pair of SMF fibers (25.2 km) are expressed in terms of the modified Allan deviation. The stabilized MCF is capable of supporting transfer of the best optical atomic clocks, despite being limited by the uncorrelated fibers from FI/FO devices. Error bars represent 1σ confidence intervals.

Using a fan-in fan-out (FI/FO) device (*36*), we launch about 2.5 mW of ultra-stable laser light into one of the cores of the 4-core MCF. At the end of 25.2 km, another FI/FO couples the light out of the MCF and into a ~ 2 m long strand of SMF-28 fiber. We use 10% of this light for the performance evaluation, while the rest is coupled back into a separate core of the MCF for the interferometric detection of the fiber noise (see Figure 1). Photodetected signals, both for performance evaluation and noise compensation, are digitally sampled, providing measurement of the frequency fluctuations and phase noise added by the MCF fiber link, as well as a digital feedback control signal, respectively.

Link performance was characterized in terms of the fractional frequency instability (*37*), shown in Fig. 3. Without active stabilization, the added fractional frequency instability from the link is $10^{-13}$ at 1 second of averaging, about three orders of magnitude above what is needed to support the transfer of state-of-the-art optical atomic clocks. Stabilization of the link reduces this instability to below $10^{-16}$ at 1 second. Even greater improvements to the link stability are realized at longer times, reaching $3\times10^{-19}$ at our longest averaging time of $10^4$ seconds. At this level, the stability of the best optical clocks can be transferred with extremely high fidelity.

Importantly, the stabilized link performance for timescales longer than 1 second is not limited by the MCF itself, but rather the standard fibers at both the transmission and receiver ends of the link, such as those that are part of the FI/FOs. We confirmed this by repeated measurements of the fractional frequency instability of the system excluding the 25 km-long MCF. In these so-called "back-to-back" measurements, we connected the FI/FO devices directly to each other, leaving the rest of the system unchanged. In this case, there remained a few spatially separated fibers where the light travels one-way, and any environmentally induced noise was much less strongly correlated among these fibers as compared to the cores of the MCF. The results for two representative back-to-back trials, shown as the gray dashed lines in Fig. 3, are comparable to the results for the full 25.2 km MCF link. Thus, while the 25.2 km link is shown to support optical clock transfer, further improvements appear attainable by, for example, simply reducing the length of the meter-long fibers used at either end of the link. Lastly, we also contrast this performance with that of a 24.4 km-long pair of standard fibers from the co-deployed SMF cable, whose performance is approximately an order of magnitude worse than a pair of cores of the MCF. This result is typical of frequency transfer over paired fibers, where each fiber is operated unidirectionally (*12*); even though the separate fibers are bundled together, their noise and instability are not sufficiently correlated to transfer signals from optical atomic clocks.

In all the measurements described above, the only light in the fiber was the ultrastable signal. Creating a multifunctional network requires compatibility with other signals within the same fiber, even within the same core. As a first step towards proving such multifunctionality, we repeated the stabilized frequency transfer over the MCF link but with simulated data traffic filling the telecom C-band, from 1530 nm to 1565 nm, with the ultrastable light multiplexed in on its own channel at 1550 nm. Data traffic was simulated with spectrally shaped broadband incoherent light co-propagating with the ultrastable laser signal, both to the remote end of the fiber and in the return path. (Using spectrally shaped amplified spontaneous emission is a proven and well-studied method to study nonlinear interference effects in fiber optic communication (*38*).) We see no degradation in the performance of the stabilized light transfer, and are again limited by the uncorrelated standard fibers at either end of the MCF. More details on this measurement are given in the Supplemental.

**Operating at the stability limit of fiber**

Since corrections to the laser frequency are only applied after the light travels a full round-trip through the fiber, every stabilized fiber link has a performance limit due to the signal propagation delay. This limit is best viewed in terms of the noise power spectral density (PSD), where the fiber noise is separated into its various frequency components. In the ideal case, where the noise between the outgoing and return signals are perfectly correlated and there is full noise cancellation on the round-trip light, the resulting frequency noise PSD at the remote end of the fiber may be expressed as (39) :

$$S_D(f) \approx a(2\pi f \tau)^2 S_{fiber}(f) \qquad \text{Eq.1.}$$

where $a$= 1/3 for uniform spatial distribution of the noise, $\tau$ is the one-way transit time over fiber, and $S_{fiber}$ is the unstabilized frequency noise PSD of the link (more details on this expression may be found in the Supplement). As shown in Fig. 4, the calculated noise limit closely follows the measured MCF stabilized link for noise offset frequencies above 1 Hz. This indicates that the link performance is as good as the best bidirectional SMF links at this length. Below 1 Hz, the predicted and measured noise curves deviate from each other due to the uncorrelated noise of the fiber pigtails. This is validated by the frequency noise of a FI/FO back-to-back trial, also shown in Fig. 4, and further confirms the stability limit imposed by these fibers.

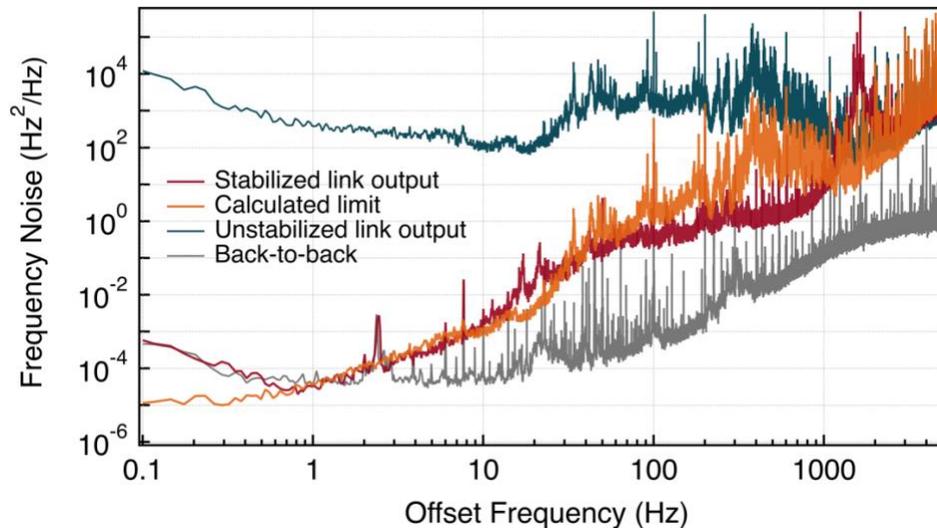

**Fig. 4. Frequency noise power spectrum**. Measured frequency noise of the link output and the calculated frequency noise limit of the stabilized link. The measurement follows the calculated limit above 1 Hz offset frequency. Below 1 Hz, the noise is limited by the uncorrelated fiber noise in the lab, which is validated by the back-to-back measurement.

### Relative timing stability

Whereas these ultrastable frequency transfer results rely on the high degree of noise correlation between cores of the MCF(40), more direct measurements on the core-to-core stability highlight the synchronization capabilities of this fiber. By launching a laser signal into all four cores simultaneously, we tracked the relative timing shifts between cores through 6.3 km of deployed MCF, the results of which are shown in Fig. 5(a). In this case, we did not stabilize the fiber link. To eliminate excess noise from the FI/FO pigtails, we used free-space optics to couple light into and out of the MCF, and extracted the relative phase shifts among the cores using digital holography (41). Details of the measurement setup may be found in the Supplemental. The relative delay between the cores is less than 10 fs (less than 2 optical cycles) over more than 7 hours of continuous measurement. This relative time delay corresponds to a fractional length difference of less than 4x10$^{-10}$, or 2.5 µm relative path length change over the duration of the measurement. With only femtosecond-level timing deviations over several hours, we cannot discount the possibility that the measurement setup itself contributes. We therefore consider this to be an upper limit on the core-to-core instability.

A statistical measure of the core-to-core timing instability is given by the time deviation, shown in Fig. 5(b). At 1 second of averaging, the core-to-core time deviation ranges from 8 to 15 attoseconds, corresponding to only a few thousandths of an optical cycle, and remains below 100 attoseconds for averaging times beyond 1000 s. For comparison, we calculated the time deviation of the unstabilized 25 km long MCF used for our frequency transfer experiments, also shown in Fig. 5(b). These results indicate that the relative core-to-core path length changes can be four orders of magnitude lower than the absolute path length change along the fiber. Such extraordinarily high relative timing stability can prove valuable in, for example, quantum networking applications such as quantum key distribution and distributed entanglement (*20*, *42*, *43*). In this case, classical light signals can reside in a separate core from the quantum channel, providing a means for extremely tight classical-quantum synchronization without the high-power classical light overwhelming the weak quantum signal.

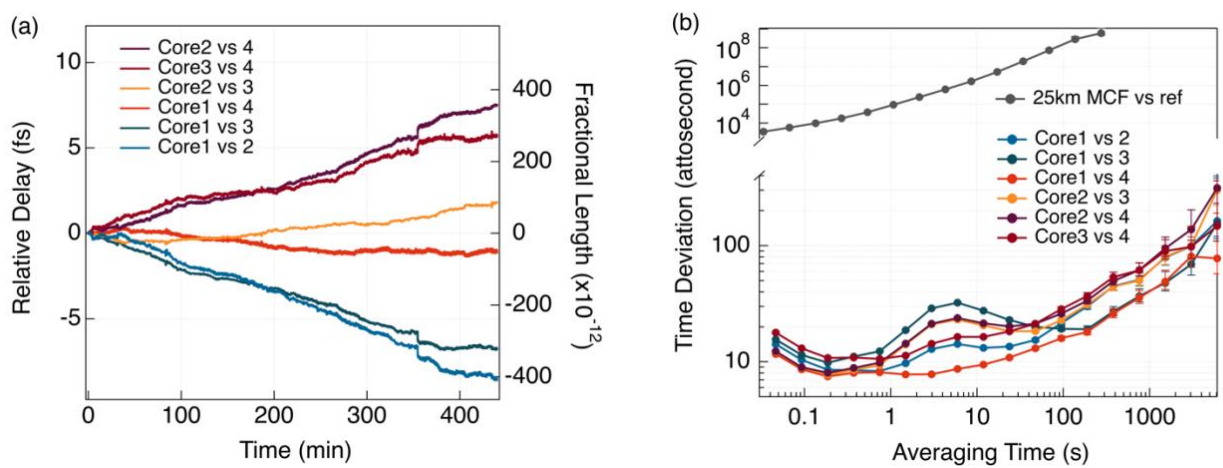

**Fig. 5. Relative timing stability among cores of the MCF**. (a) Core-to-core relative delay and fractional length changes of the unstabilized 6.3 km deployed MCF (no FI/FO) over >7 hours. The time delay between cores stays below 10 fs, equivalent to <0.4 part per billion fractional length change over the experiment duration. The first 80 minutes of data affected by human activity near the measurement setup is removed. See Supplemental for the complete dataset. (b) Time deviation (TDEV) between cores of the unstabilized 6.3 km-long deployed MCF. TDEV of one core of the 25km-long unstabilized MCF relative to a reference laser is included for comparison.

Multicore optical fibers are well-positioned to be the future of the long-haul fiber optic networks. We show ultrastable frequency transfer through 25.2 km-long deployed MCF with fractional instability reaching $3\times10^{-19}$, obtained while co-propagating with other optical signals and without bi-directional use of any one core. This demonstration shows the compatibility of in-network MCF for optical clock comparisons, expanding the application space of these optical fibers beyond data transfer. Longer transfer distances will require the integration of multicore optical amplifiers (*44*), with the ultimate goal of intercontinental clock comparisons with precision that is otherwise unattainable. As MCF optical amplifier testbed demonstrations have already established the capability of low noise, high-fidelity data transfer over 1000s of km (*31*), we see no barrier to the stabilization of a long distance, amplified MCF link. Additionally, the exquisite relative timing stability between cores can enable quantum networking modalities with the highest synchronization requirements. With the wide-scale deployment of MCF on the horizon, a vast network serving a multiplicity of scientific and technological applications is within reach.

**Acknowledgment**
We thank A. Ludlow, F. Giorgetta and E. Donley for their comments on this manuscript.
**Funding:** This work is supported by National Institute of Standards and Technology and Project INCIPICT (Innovating city Planning through Information and Communications Technologies).

**Authors contributions:** NH and FQ conceptualized and designed the fiber link stabilization experiment. NF, MM and CA conceptualized and designed the digital holography experiment. NH, FQ, NF, MM, GDS, DS and CA conducted the multicore fiber stabilization experiments. NF and MM conducted the digital holography experiments. YL, TN, DL, CM built and supported the compact ultrastable cavity stabilized laser system. FQ, NF, TH, AM and CA funding acquisition. TH, AM, and CA fiber testbed deployment. NH and FQ wrote the manuscript with input from all authors.

**Competing interests:** TH is an employee of Sumitomo Electric a supplier of multicore fiber such as the one in used in this study.

**Data and materials availability:** The data from the main text and supplementary materials are available from the NIST Public Data Repository. This is a contribution of the National Institute of Standards and Technology, not subject to U.S. copyright.

# Supplemental material

## Fiber optic test bed in L'Aquila

We performed the measurements presented in this paper at a fiber optic testbed located in the city of L'Aquila, Italy. The test bed includes a bundle of SMF28 and 3 different types of MCF; 4 strands of 4-core uncoupled MCF each 6.29 km long, 12 strands of 4-core coupled MCF each 6.29 km long and, one strand of 8-core uncoupled core MCF ~69.2 km long (14). We used the 4-core uncoupled MCF for this work. The fiber cladding diameter is 125 µm and the mode field diameter of each core is 8.4-8.5 µm at wavelength of 1310 nm, which are consistent with ITU-T G.652 recommendations. This fiber is the only uncoupled MCF in the tunnel that covers the C-band. Hence, we chose this fiber over the 8-core uncoupled MCF which supports only O-band. The wavelength coverage of the 4-core uncoupled MCF is from O to L-band and has a loss of ~ 0.2 dB/km. The cables containing the MCFs are deployed in a tunnel under the city, with ~0.66 km of the fiber in the room where the fiber noise stabilization equipment was installed, providing access to both ends of the fiber. In the room, the Fan-In Fan-Out devices with 2m-long pigtails that couple to the MCF and are mounted on a 19-inch rack. More details on the MCF testbed can be found in (14).

## Cavity-stabilized light source

An important consideration for the light source used for optical frequency transfer experiment is the laser's phase noise. The laser's noise should be lower than the fiber link that is used for transferring the light, otherwise the stabilization servo measures the laser noise rather than that of the fiber link. For this reason, we use a transportable cavity-stabilized laser for our experiment. The light from a fiber laser centered at 1550 nm is stabilized to a compact, vacuum gap, high-finesse cavity with the finesse near 550,000. The cylindrical cavity is 19.05 mm long and 25.4 mm in diameter (9.7 mL volume) (34). In addition, this compact cavity is bonded in vacuum and does not require a vacuum enclosure or vacuum pump for operation which makes it ideal for operating in the field. We lock the laser to this cavity using Pound-Drever-Hall (PDH) locking technique. The phase noise of the cavity-stabilized light is thermal noise-limited up to 10 kHz offset and exhibits $10^{-14}$ level fractional frequency instability at 1 second of averaging.

## Details of the experimental setup

A detailed schematic of the experimental setup is shown in Fig. S1. The laser light interferometer consists of a short arm, indicated as the reference arm in the figure, where light is retro-reflected using a free-space mirror. The light in the interferometer's long arm is frequency shifted by 100 MHz using an acousto-optic modulator (AOM). After travelling back over 25.2 km of MCF, the light passes again through the AOM, then combines with the light from the reference arm. The heterodyne beat note between reference light and returned and twice shifted light at 200 MHz is detected using a fast photodetector. The RF beat note is amplified and its frequency is divided by 8 and used as the input to a digital servo box with computer-controlled user interface. The servo applies correction to the transmitted light by acting on the frequency of light by changing the tuning voltage of the voltage-controlled oscillator that drives the AOM.

We use all-digital servo and data acquisition (DAQ) systems in our experiment. The field programable gate array (FPGA)-based DAQ samples the in-loop and out-of-loop (output of the link) signals simultaneously. The signals are initially sampled at the rate of 2.5GS/s. Since this data rate for two simultaneous channels is above the capacity of the 1G Ethernet cables, we reduce the sample rate before transferring data over Ethernet cable and saving the data on a hard drive. The process is as follows: The sampled data for each channel is lowpass filtered and down-sampled by factor of 10. We then apply a moving average over the down-sampled data (window size is 1024) and down-sample the data again. The final sample rate of the recorded data is 244 kS/s. The phase and the frequency data are extracted in post processing. We use python package AllanTools for calculating MDEV and TDEV presented in this paper.

The servo (also FPGA-based) and DAQ are externally clocked with a common 10 MHz signal.

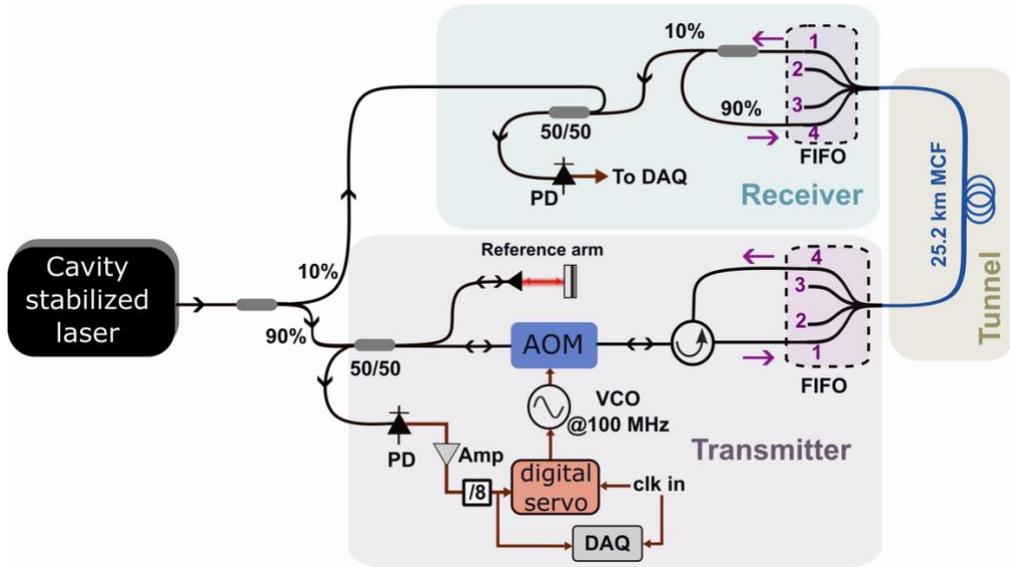

**Figure S1.** Detailed schematic of the stabilized MCF link. AOM: acousto-optic modulator, FIFO: fan in-fan out device, PD: photodetector, VCO: voltage-controlled oscillator, DAQ: data acquisition system, clk in: clock in (10 MHz).

The optical power launched into core1 of the 4-core MCF (measured before FIFO) is 2.5 mW. The received power at the end of the 25.2 km MCF is 0.25 mW, from which 90% is sent back over core4 of the fiber. At the transmitter side, we measure ~22 µW of return light before the AOM. The roundtrip loss through the MCF link, including the loss of FI/FO devices, is ~20 dB.

We studied link stabilization using different core combinations, i.e. cores next to each other and cores across from each other. We do not observe any significant changes in the performance of the stabilized link using different core combinations for stabilization. As discussed in the main text, the performance of the link for timescales longer than ~1 second is limited by the noise of the uncorrelated fibers, as confirmed by back-to-back trials.

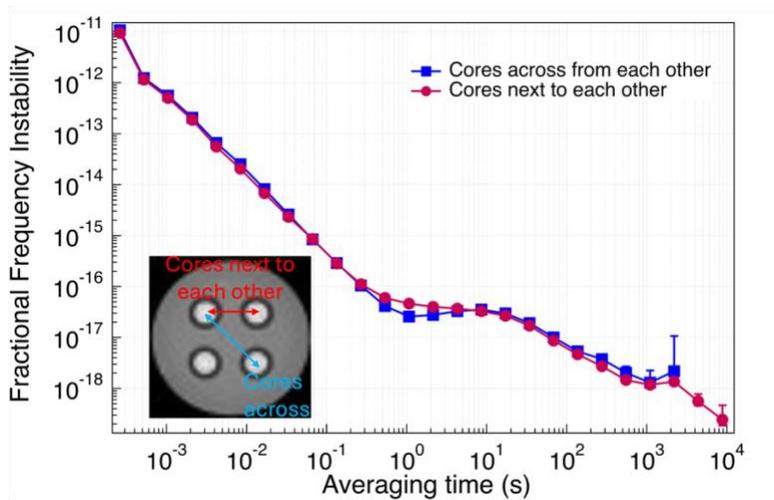

**Figure S2.** Fractional frequency instability, expressed in terms of the modified Allan deviation, of stabilized 25.2 km MCF link. Different core combination used for stabilization show no difference in performance.

In addition to the 25.2 km long link, we also stabilized a 6.3 km section of the MCF link using the same stabilization method. The fractional frequency instability of the 6.3 km MCF is shown below (Fig. S3). The data presented here for the 6.3 km section is collected with a frequency counter with 1s gate time. Again, the measured stability is limited by the SMF fibers in the lab.

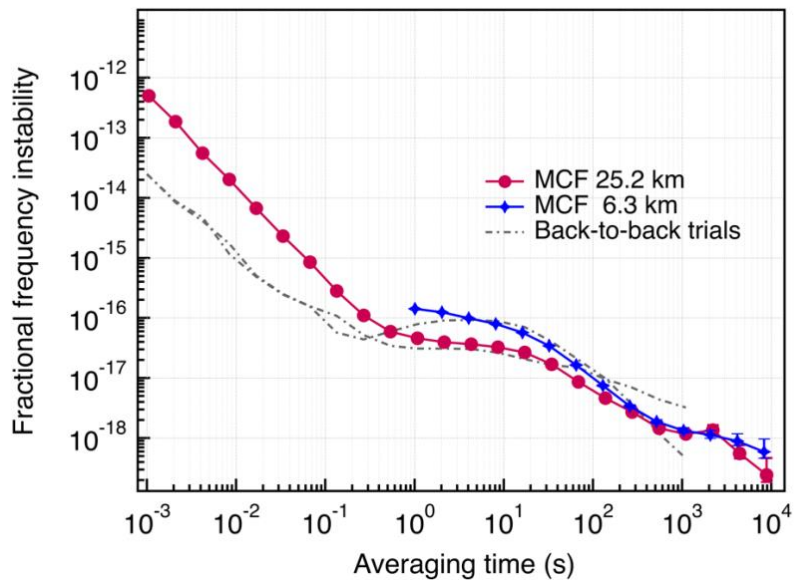

**Figure S3.** Fractional frequency instability of stabilized 6.3 km section of MCF. The 6.3 km performance follows the 25.2 km fiber indicating that increasing the fiber length did not affect the performance of the stabilized link.

## Stabilized MCF link with light in other wavelength channels

As discussed in the main text, in addition to only transmitting the ultrastable light over MCF, we studied the stability of the transferred light while all other telecommunications channels in the C-band were occupied by light from a broadband light source. The figure below shows the optical spectrum of the broadband light after multiplexing with the ultrastable light at 1550 nm. The broadband light source is amplified in two stages and sent through a programmable optical filter, where light is dispersed onto a liquid crystal-on-silicon array to provide amplitude shaping. With this filter, we created a 100 GHz wide notch in the spectrum at 1550.12 nm (193.4 THz), corresponding to the stable laser wavelength. The light from the cavity stabilized laser is then wavelength-multiplexed with the broadband light and sent over the MCF. The total optical power of the broadband light is 36 mW. The inset shows the zoomed-in optical spectrum around the ultrastable light.

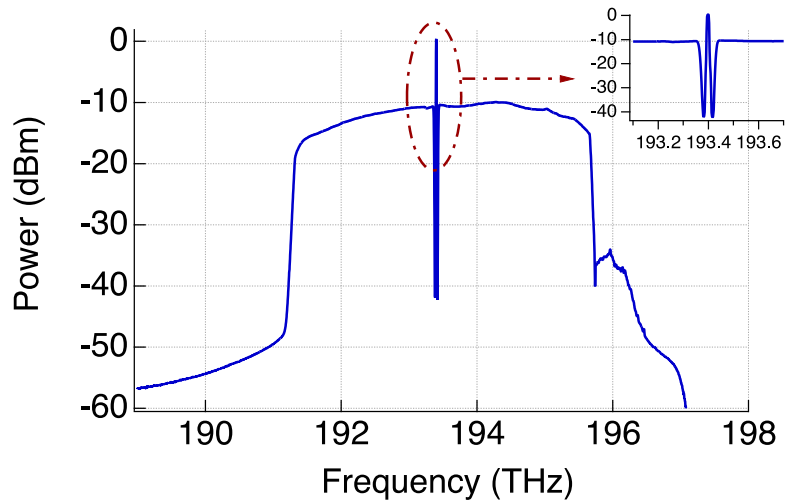

**Figure S4**. Optical spectrum of the broadband light covering C-band and multiplexed with ultrastable laser light. The inset is the zoomed-in view of the ultrastable laser inserted in the telecommunications channel 34.

## Frequency noise

For evaluating the stabilized link performance, we measured the power spectral density of the unstabilized and stabilized link output. As seen in Fig. S5, we are able to suppress the noise of the 25.2 km-long unstabilized MCF link by ~ 72 dB at 1 Hz frequency offset. The lock bandwidth, as discussed in the main text, is limited by the transit time of the light through the fiber. From Fig. S5, the lock bandwidth for the 25km-long link is 1.6 kHz, close to the theoretical maximum of ~ 1.9 kHz. This limit is calculated from the one-way transit time over the fiber of 0.126 ms. In addition to frequency noise of the link output, Fig. S5 shows the in-loop residual noise. The fact that the in-loop noise is lower than the stabilized output noise is consistent with the link being delay limited, as given in the main text by Eq. 1.

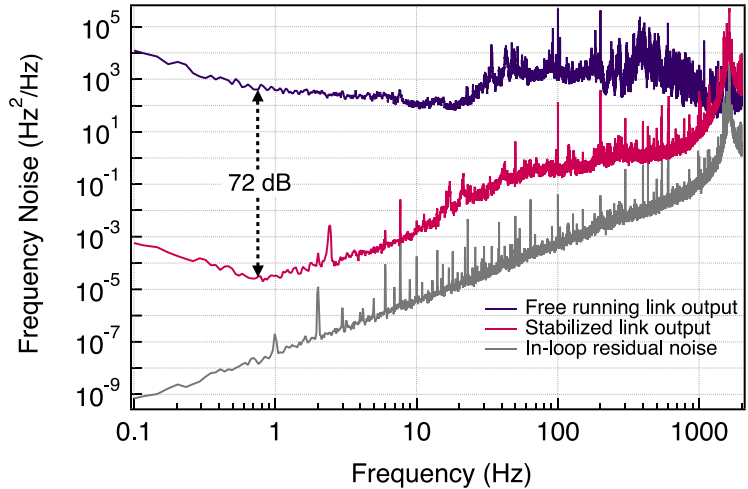

**Figure S5.** Frequency noise of the MCF link output. Noise suppression of 72 dB is observed at 1 Hz offset frequency.

## Digital holography measurement

Digital holography is a straightforward technique for making a multi-channel phase detector and avoiding all fiber pigails. The beams travel together within 1x1mm area through a small section of free space. The closeness of the beams through the free-space section minimizes any non-correlated phase in this section. A Gaussian reference beam from a SMF is interfered with the beams exiting MCF and recorded on a fast camera (2000 fps). The MCF facet is imaged on the camera via a 4-f telescope. After the telescope, a polarization beam splitter spatially separated the polarizations producing 8 beams on the camera corresponding to the output vertical and horizontal polarization. A photo of the experimental setup and an example of recorded interference fringes is shown in Fig S6. Complete detail of the setup is given in (39).

The phase and intensity of each core is extracted from the interference pattern by digital holography. The steps are as follows: 1) 2D fast Fourier transform (FFT) of the interference pattern to convert space to spatial frequency, 2) filtering out the signal at the fringe spatial frequency, 3) inverse 2D FFT to return to the spatial domain, 4) spatial selection of each core, 5) removing any residual linear and quadratic phase, 6) computing the average phase via summation of each core. Then these phases can all be compared.

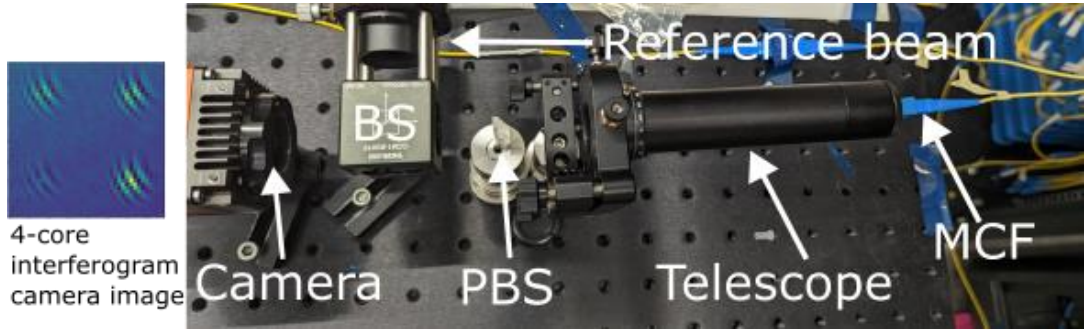

**Figure S6.** Free-space digital holography setup for measuring all four cores in parallel without the use of fiber pigtails. Raw camera image of the received X-polarization showing the interference fringes. BS: non-polarizing beam splitter; PBS: polarizing beam splitter.

The complete measured core-to-core time delay is shown in Figure S7. The spikes in the dataset in data are due to human activities around the setup.

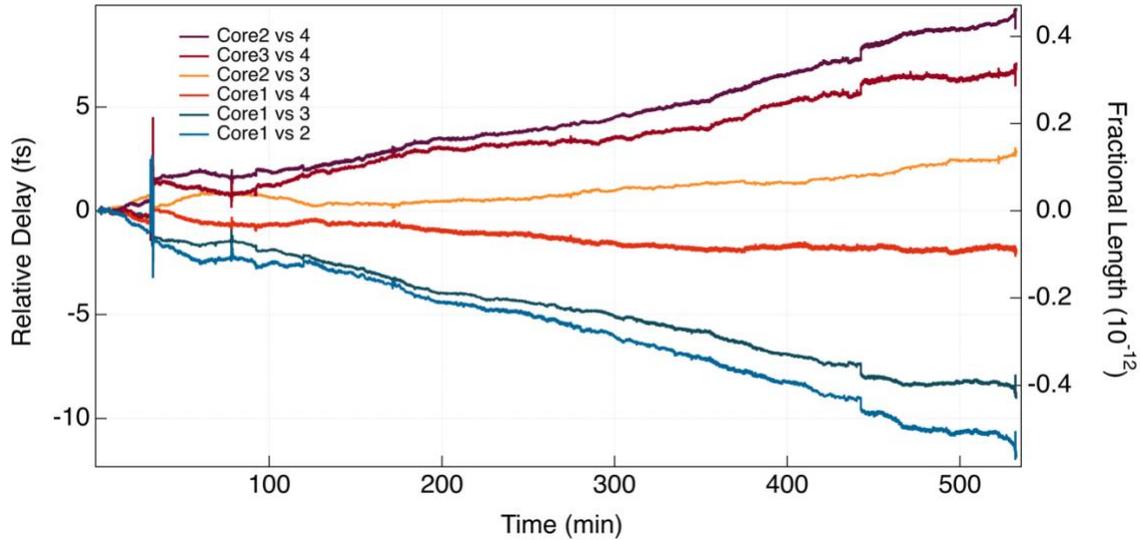

**Figure S7.** Core-to-core time delay (without FI/FO) measurement using the digital holography setup.

## Contribution to the phase noise spectrum of the noise of the FIFO pigtails

Unlike bidirectional SMF, the use of MCF for ultrastable frequency transfer necessitates finite lengths of uncorrelated fibers. In this section, we extend the results of (5) to include uncorrelated fibers at the transmission and remote ends. Let $\varphi_{\text{fiber}}(t)$ be the phase noise at the remote output of the fiber:

$$\varphi_{\text{fiber}}(t) = \varphi_{in,+}(t - \tau) + \int_0^L \delta\varphi_z(z, t - (\tau - z/c)) \mathrm{d}z + \varphi_{out,+}(t),$$

where $\delta\varphi_z$ is the phase noise added per unit length by the fiber in units of rad/m, $\tau = L/c$ is the delay over length $L$ with $c = c_0/n$ the speed of light in a medium with refractive index $n$, $\varphi_{in,+}$ is the phase noise added by the fan-in at input, and $\varphi_{out,+}$ is the phase noise added by the fan-out at the remote end. The phase noise after the roundtrip is

$$\varphi_{\text{fiber,RT}}(t) = \varphi_{in,+}(t - 2\tau) + \int_0^L \delta\varphi_z\big(z, t - (2\tau - z/c)\big)dz + \varphi_{out,+}(t - \tau) + \varphi_{in,-}(t - \tau)$$

$$+ \int_0^L \delta\varphi_z(z, t - z/c)dz + \varphi_{out,-}(t),$$

where $\varphi_{in,-}$ is the noise added by the fan-in at the remote end and $\varphi_{out,-}$ is the phase noise added by the fan-out at the near end, both belonging to the return path. Using for the Fourier transform a definition with a plus sign in front of the imaginary unit in the exponential (a convention opposite to that followed in the appendix of *(5)*):

$$\tilde{f}(\omega) = \int_{-\infty}^{\infty} f(t) e^{i\omega t} dt,$$

we obtain:

$$\tilde{\varphi}_{\text{fiber}}(\omega) = e^{i\omega\tau} \tilde{\varphi}_{in,+}(\omega) + e^{i\omega\tau} \int_0^L e^{-i\omega z/c} \delta\tilde{\varphi}_z(z, \omega) dz + \tilde{\varphi}_{out,+}(\omega),$$

$$\tilde{\varphi}_{\text{fiber,RT}}(\omega) = 2e^{i\omega\tau} \int_0^L \cos\big(\omega(\tau - z/c)\big) \delta\tilde{\varphi}_z(z, \omega) dz + e^{2i\omega\tau} \tilde{\varphi}_{in,+}(\omega)$$

$$+ e^{i\omega\tau}\big[\tilde{\varphi}_{out,+}(\omega) + \tilde{\varphi}_{in,-}(\omega)\big] + \tilde{\varphi}_{out,-}(\omega).$$

With the feedback loop turned on, the spectrum of the phase fluctuations at the remote end is *(5)* :

$$\tilde{\varphi}_{on}(\omega) = \tilde{\varphi}_{\text{fiber}}(\omega) - e^{i\omega\tau} \frac{G(\omega)}{1 + G(\omega)} \frac{\tilde{\varphi}_{\text{fiber,RT}}(\omega)}{1 + e^{2i\omega\tau}},$$

where $G(\omega) = G_0 F(\omega)(-i\omega)^{-1}\big(1 + e^{2i\omega\tau}\big)$ is the loop gain, with $G_0$ a cumulative loop gain in units of Hz/rad which includes the response of the phase detector (in V/rad) and the transfer function of the voltage controlled oscillator (in Hz/V), and $F(\omega)$ the dimensionless loop filter gain. For $|G(\omega)| \to \infty$ we have:

$$\tilde{\varphi}_{on}(\omega) = \tilde{\varphi}_{\text{fiber}}(\omega) - \frac{\tilde{\varphi}_{\text{fiber,RT}}(\omega)}{2\cos(\omega\tau)}.$$

Notice that the divergence for $\omega\tau = \pi/2$, that is for $f = 1/(4\tau)$, is apparent, because $G(\omega) = e^{i\omega\tau} G_0 F(\omega)(-i\omega)^{-1} 2\cos(\omega\tau)$ also vanishes. After some algebra we obtain:

$$\tilde{\varphi}_{\text{on}}(\omega) = \tilde{\varphi}_{\text{on,fiber}}(\omega) + \tilde{\varphi}_{\text{on,FIFO}}(\omega),$$

where we have separated the contribution of the fiber

$$\tilde{\varphi}_{\text{on,fiber}}(\omega) = -e^{-i\omega\tau}[i + \tan(\omega\tau)] \int_0^L \sin(\omega z/c)\, \delta\tilde{\varphi}_z(z,\omega)\mathrm{d}z,$$

from that of the FIFO pigtails

$$\tilde{\varphi}_{\text{on,FIFO}}(\omega) = \frac{1}{2\cos(\omega\tau)}\left[\tilde{\varphi}_{in,+}(\omega) + e^{-i\omega\tau}\tilde{\varphi}_{out,+}(\omega) - e^{i\omega\tau}\tilde{\varphi}_{in,-}(\omega) - \tilde{\varphi}_{out,-}(\omega)\right].$$

Notice that if the noise of the pigtails of the FIFO at input and at the remote end are equal, that is if $\tilde{\varphi}_{in,+}(\omega) = \tilde{\varphi}_{out,-}(\omega)$, and $\tilde{\varphi}_{out,+}(\omega) = \tilde{\varphi}_{in,-}(\omega)$, then the contribution to the noise of the FIFO at input vanishes and one gets $\tilde{\varphi}_{\text{FIFO}}(\omega) = i\tan(\omega\tau)\tilde{\varphi}_{out,+}(\omega)$. As a result, the noise contributions of the FIFO pigtails vanish for $\omega\tau \to 0$.

Let us now consider the realistic case in which the phase noise contributions of the pigtails are independent of each other and independent of the fiber fluctuations. Under this assumption, we have

$$\langle|\tilde{\varphi}_{\text{on}}(\omega)|^2\rangle = \langle|\tilde{\varphi}_{\text{on,fiber}}(\omega)|^2\rangle + \langle|\tilde{\varphi}_{\text{on,FIFO}}(\omega)|^2\rangle.$$

Let us consider the contribution of the FIFO pigtails first. We have

$$\langle|\tilde{\varphi}_{\text{FIFO}}(\omega)|^2\rangle = \frac{1}{4\cos^2(\omega\tau)} S_{\text{FIFO}}(\omega),$$

with

$$S_{\text{FIFO}}(\omega) = \left[\langle|\tilde{\varphi}_{in,+}(\omega)|^2\rangle + \langle|\tilde{\varphi}_{out,+}(\omega)|^2\rangle + \langle|\tilde{\varphi}_{in,-}(\omega)|^2\rangle + \langle|\tilde{\varphi}_{out,-}(\omega)|^2\rangle\right]$$

the sum of the power spectra density of the phase noise added by the FIFO pigtails.

Consider now the contribution of the fiber. We have

$$\langle|\tilde{\varphi}_{\text{on,fiber}}(\omega)|^2\rangle = \frac{1}{\cos^2(\omega\tau)} \int_0^L \mathrm{d}z \int_0^L \mathrm{d}z'\, \sin\left(\frac{\omega z}{c}\right) \sin\left(\frac{\omega z'}{c}\right) \langle\delta\tilde{\varphi}_z^*(z,\omega)\delta\tilde{\varphi}_z(z',\omega)\rangle,$$

where the star denotes complex conjugation. Assume that the phase noise added by the fiber has a finite correlation length $L_{\text{corr}}$, with spatial correlation function

$$\langle \delta\tilde{\varphi}_z^*(z,\omega)\delta\tilde{\varphi}_z(z',\omega)\rangle = C(|z-z'|)\,\langle|\delta\tilde{\varphi}_z(\omega)|^2\rangle,$$

where $C(|z-z'|)$ is a dimensionless smooth function that goes to zero sufficiently fast for $|z-z'| \to \infty$, such that $C(0) = 1$ and

$$\int_{-\infty}^{\infty} C(|z-z'|)\mathrm{d}z' = L_{\text{corr}}.$$

A possible choice for $C(|z-z'|)$ is a two-sided exponential distribution, $C(|z-z'|) = e^{-2|z-z'|/L_{\text{corr}}}$. If we now assume $L_{\text{corr}} \ll L$ and $L_{\text{corr}} \ll c/\omega$, the term $L_{\text{corr}}^{-1} C(|z-z'|)$ can be replaced in Eq. (1) by a Dirac delta function, so that Eq. (1) yields

$$\langle|\tilde{\varphi}_{\text{on,fiber}}(\omega)|^2\rangle = \frac{L_{\text{corr}}}{\cos^2(\omega\tau)} \int_0^L \sin^2\left(\frac{\omega z}{c}\right) \langle|\delta\tilde{\varphi}_z(\omega)|^2\rangle \, \mathrm{d}z,$$

that is, after integration over $z$

$$\langle|\tilde{\varphi}_{\text{on,fiber}}(\omega)|^2\rangle = \frac{1 - \text{sinc}(2\omega\tau)}{2\cos^2(\omega\tau)} \langle|\delta\tilde{\varphi}_z(\omega)|^2\rangle \, L \, L_{\text{corr}}.$$

Using the same approximation on $\langle|\tilde{\varphi}_{\text{fiber}}(\omega)|^2\rangle$ we obtain

$$\langle|\tilde{\varphi}_{\text{fiber}}(\omega)|^2\rangle = \langle|\tilde{\varphi}_{in,+}(\omega)|^2\rangle + \langle|\tilde{\varphi}_{out,+}(\omega)|^2\rangle + \langle|\delta\tilde{\varphi}_z(\omega)|^2\rangle \, L \, L_{\text{corr}},$$

so that

$$\langle|\tilde{\varphi}_{\text{on,fiber}}(\omega)|^2\rangle = \frac{1 - \text{sinc}(2\omega\tau)}{2\cos^2(\omega\tau)} \left[\langle|\tilde{\varphi}_{\text{fiber}}(\omega)|^2\rangle - \langle|\tilde{\varphi}_{in,+}(\omega)|^2\rangle - \langle|\tilde{\varphi}_{out,+}(\omega)|^2\rangle\right].$$

Using now that $\langle|\tilde{\varphi}_{\text{fiber}}(\omega)|^2\rangle \gg \langle|\tilde{\varphi}_{in,+}(\omega)|^2\rangle + \langle|\tilde{\varphi}_{out,+}(\omega)|^2\rangle$, we obtain the expression

$$\langle|\tilde{\varphi}_{\text{on}}(\omega)|^2\rangle = \frac{1}{4\cos^2(\omega\tau)} \left[2\bigl(1 - \text{sinc}(2\omega\tau)\bigr) \langle|\tilde{\varphi}_{\text{fiber}}(\omega)|^2\rangle + S_{\text{FIFO}}(\omega)\right].$$

A measurement of the phase noise $\langle|\tilde{\varphi}_{\text{on}}(\omega)|^2\rangle$ in a back-to-back measurement, which corresponds to $\tau = 0$, allows one to extract $S_{\text{FIFO}}(\omega)$, the phase noise added by the FIFO pigtails. Then, a measurement of the phase noise at the remote end $\langle|\tilde{\varphi}_{\text{on}}(\omega)|^2\rangle$ allows one to measure $\langle|\tilde{\varphi}_{\text{fiber}}(\omega)|^2\rangle$, the phase noise added by the fiber.

For small $\omega\tau$, neglecting terms of the order of $(\omega\tau)^4$ and larger, we obtain

$$\langle |\tilde{\varphi}_{\text{on}}(\omega)|^2 \rangle = \frac{1}{4} S_{\text{FIFO}}(\omega) + \frac{1}{3}(\omega\tau)^2 \langle |\tilde{\varphi}_{\text{fiber}}(\omega)|^2 \rangle + O[(\omega\tau)^4].$$

Changing notation to match the main text, where $\langle |\tilde{\varphi}_{\text{on}}(\omega)|^2 \rangle \rightarrow S_D(\omega)$ and $\langle |\tilde{\varphi}_{\text{fiber}}(\omega)|^2 \rangle \rightarrow S_{fiber}(\omega)$, this result reduces to Eq. 1 when the contribution from the FIFOs is ignored. For small $\omega\tau$, the dominant noise contribution to the phase noise spectrum is the FIFO pigtails when $S_{\text{FIFO}}(\omega) > (\omega\tau)^2 S_{fiber}(\omega)$, a condition met in our experiments for offset frequencies less than ~1 Hz.